\documentclass[twocolumn,tighten,times]{aastex62}

\hypersetup{linkcolor=black,citecolor=black,filecolor=cyan,urlcolor=magenta}

\newcommand{\um}{$\mu$m}

\accepted{ for Publication to ApJ}

\shorttitle{ALMA polarimetry of NGC 1068}
\shortauthors{Lopez-Rodriguez et al.}

\begin{document}

\title{ALMA polarimetry measures magnetically aligned dust grains in the torus of NGC 1068}

\correspondingauthor{Lopez-Rodriguez, E.}
\email{enrique.lopez-rodriguez@nasa.gov}

\author{Enrique Lopez-Rodriguez}
\affil{SOFIA Science Center, NASA Ames Research Center, Moffett Field, CA 94035, USA}

\author{Almudena Alonso-Herrero}
\affil{Centro de Astrobiolog\'ia (CSIC-INTA), ESAC Campus, E-28692 Villanueva de la Canada, Madrid, Spain}

\author{Santiago Garc\'ia-Burillo}
\affil{Observatorio Astron\'omico Nacional (OAN-IGN)-Observatorio de Madrid, Alfonso XII, 3, 28014, Madrid, Spain}

\author{Michael S. Gordon}
\affil{SOFIA Science Center, NASA Ames Research Center, Moffett Field, CA 94035, USA}

\author{Kohei Ichikawa}
\affil{Frontier Research Institute for Interdisciplinary Sciences, Tohoku University, Sendai 980-8578, Japan}
\affil{Astronomical Institute, Tohoku University, Aramaki, Aoba-ku, Sendai, Miyagi 980-8578, Japan}

\author{Masatoshi Imanishi}
\affil{National Astronomical Observatory of Japan, Mitaka, Tokyo 181-8588, Japan}

\author{Seiji Kameno}
\affil{National Astronomical Observatory of Japan, Mitaka, Tokyo 181-8588, Japan}
\affil{Nobeyama Radio Observatory, National Astronomical Observatory of Japan 462-2 Nobeyama, Minamisaku, Nagano 384-1305, Japan}

\author{Nancy A. Levenson}
\affil{Space Telescope Science Institute, 3700 San Martin Dr, Baltimore, MD 21218, USA}

\author{Robert Nikutta}
\affil{1National Optical Astronomy Observatory, 950 N Cherry Ave, Tucson, AZ 85719, USA}

\author{Chris Packham}
\affil{The University of Texas at San Antonio, One UTSA Circle, San Antonio, TX 78249, USA}



\begin{abstract}

The obscuring structure surrounding active galactic nuclei (AGN) can be explained as a dust and gas flow cycle that fundamentally connects the AGN with their host galaxies. This structure is  believed to be associated with dusty winds driven by radiation pressure. However, the role of magnetic fields, which are invoked in almost all models for accretion onto a supermassive black hole and outflows, is not thoroughly studied. Here we report the first detection of polarized thermal emission by means of magnetically aligned dust grains in the dusty torus of NGC 1068 using ALMA Cycle 4 polarimetric dust continuum observations ($0\farcs07$, $4.2$ pc; 348.5 GHz, $860$ \um). The polarized torus has an asymmetric variation across the equatorial axis with a peak polarization of $3.7\pm0.5$\% and position angle of $109\pm2^{\circ}$ (B-vector) at $\sim8$ pc east from the core. We compute synthetic polarimetric observations of magnetically aligned dust grains assuming a toroidal magnetic field and homogeneous grain alignment. We conclude that the measured 860 \um\ continuum polarization arises from magnetically aligned dust grains in an optically thin region of the torus. The asymmetric polarization across the equatorial axis of the torus arises from 1) an inhomogeneous optical depth, and 2) a variation of the velocity dispersion, i.e. variation of the magnetic field turbulence at sub-pc scales, from the eastern to the western region of the torus. These observations and modeling constrain the torus properties beyond spectral energy distribution results. This study strongly supports that magnetic fields up to a few pc contribute to the accretion flow onto the active nuclei.

\end{abstract}




\section{Introduction} \label{sec:int}

 The interface between the active galactic nuclei (AGN) and their host galaxies is a region of a few pc in size with a gas and dust flow cycle, a so-called torus \citep[e.g.][]{KB1988,SKB2011,DBKK2011,wada2012}. The torus and accretion disk fuel the accretion onto the supermassive black hole (SMBH) and launch outflows, which fundamentally connect the black holes to their host galaxies. This region is particularly interesting for several reasons: 1) from infrared (IR) to X-ray, aside from synchrotron emission, the total flux spectral energy distributions (SEDs) of radio-loud and radio-quiet AGN are virtually identical \citep[e.g.][]{mullaney2011, HA2018}, and 2) advected magnetic fields are strongly implicated in almost all models for the launching and collimations of jets, accretion onto the SMBH, and outflows \citep[e.g.][]{BZ1977,BP1982,ghisellini2014}. The magnetic fields allow for both the transport of angular momentum in the disk from sub-pc scales to the black hole and the formation of jets. Although the torus represents a section of the AGN accretion flow on pc scales, the role of magnetic fields in this pc-scale accretion flow is poorly understood.

The potential influence by the magnetic field of the AGN on the dusty torus and accretion activity can be investigated using the polarization signature of magnetically aligned dust grains. Dust grains can be aligned by the presence of magnetic fields described by theories of radiative torques \citep[RATs][]{HL2009} and also by intense radiation fields or outflowing media. As radiation propagates through these aligned and elongated dust grains, preferential extinction of radiation along one plane leads to a measurable polarization in the transmission of this radiation--a term called dichroic absorption. The short axis of the dust grains aligns with the local magnetic field, and the observed position angle (PA) of the polarization traces the direction of the magnetic field. An observed PA of polarization perpendicular to the magnetic field is expected for polarization in emission. However, optical depth effects, turbulent media, and spatial resolution of the observations make the interpretation of magnetically aligned dust grains very complex. In conjunction with high-angular resolution observations, polarization studies, where optically thin dust emission dominates, are an excellent tool for characterizing magnetic fields in the torus.

With more than 52 years of polarimetric observations at all wavelengths and spatial resolutions \citep{marin2018}, NGC 1068 \citep[D = 14.4 Mpc][and $1$\arcsec = $60$ pc, adopting $H_{0}$ = 73 km s$^{-1}$ Mpc$^{-1}$]{BH1997} is the most studied AGN. From ultraviolet (UV) to optical wavelengths, the polarized core is dominated by dust scattering around the AGN \citep{MA1983,AHM1994}. At near-IR (NIR), the polarization arises from dichroic absorption by aligned dust grains within a $\sim0\farcs3$ (18 pc) unresolved core \citep{bailey1988,young1995,packham1997,simpson2002,LR2015,gratadour2015}. At mid-IR (MIR), the  $\sim0\farcs3$ (18 pc) unresolved core is measured to be unpolarized \citep{packham2007}, which is attributed to self-absorbed polarized dust emission by aligned dust grains within the torus \citep{LR2016}. Although these studies suggest the presence of aligned dust grains in the torus, the core remains unresolved at these wavelengths and only the integrated signature of the torus is studied. ALMA is currently the only facility that can resolve the torus in dust continuum emission and molecular lines. Recent ALMA observations \citep{GB2016,Imanishi2016,Imanishi2018} have observed that the torus of NGC 1068 is a highly inhomogeneous molecular disk with a size of $12\times5$ pc ($0\farcs2 \times 0\farcs08$). This structure is rotating with the eastern side redshifted and the western side blueshifted, where the western side has large velocity dispersion and strong emission in the HCN J=3-2 and HCO$^{+}$ J = 3-2 lines. At sub-mm wavelengths, the dust is optically thin and the bulk emission of the dusty disk is well sampled. Therefore, sub-mm polarimetry will reveal a clearer picture of the presence of large-scale magnetic fields in the torus than at shorter wavelengths.
 
 Our goal is to obtain the polarization signature of magnetically aligned dust grains using resolved images of the dust continuum emission of the torus of NGC 1068. We can achieve this goal using ALMA high-angular resolution ($0\farcs07$, 4.2 pc) polarimetric observations at Band 7 ($348.5$ GHz, $860$ \um) of continuum emission of the nucleus of NGC 1068. Here we present the resolved polarized emission of the torus and model the expected polarization signature by magnetically aligned dust grains. The paper is organized as follows: Section \ref{sec:obs} describes the observations and data reduction. Observational results are shown in Section \ref{sec:results}. Section 4 presents our magnetic field model of the torus, which is then analyzed and discussed in Section \ref{sec:DIS}. In Section \ref{sec:CON}, we present the conclusions.


\section{Observations and data Reduction} \label{sec:obs}

We performed (ID: 2016.1.00176.S, PI: Lopez-Rodriguez) Cycle 4 ALMA band 7 (348.5 GHz, 4.78 GHz bandwidth) observations of NGC 1068 using the full polarization mode centered at 02h42m40.80s, -00$^{\circ}$00\arcmin47.8\arcsec\ ICRS. Our observations were performed using the C40-7 nominal configuration, which provided an angular resolution of $0.077\arcsec \times 0.058\arcsec$ with a position angle of 75.13$^{\circ}$ and execution time of 550.2 minutes on 20170824/25. The largest recoverable scale in the data is approximately $0\farcs9$, which is larger than the expected extension (i.e. diameter) of the torus estimated to be $12 \times 5$ pc ($0\farcs2~\times $ 0\farcs08) \citep{Imanishi2018}. Our observations comprise 8.0 GHz of wide-band dust continuum ranging in the $336.495 - 350.495$ GHz centered at 348.5 GHz (860 \um). The flux and bandpass calibrator was J0006-0623, and phase calibrator was J0239-0234; these calibrators were chosen automatically by querying the ALMA source catalog when the project was executed. The polarization calibrator, J0006-0623, was chosen because of its high and stable polarization fraction over a period of a few months prior to these observations. ALMA's flux calibration is $\sim10$\%, as determined by their flux monitoring program. The uncertainties estimated in this work are all statistically based on aperture photometry on the area of interest in the object.

The dust continuum and polarization images were estimated using the Common Astronomy Software Applications \citep[\textsc{casa}, ][]{mcmullin2007} with the task \textsc{tclean} with a natural weighting parameter. The rms noise level of the final Stokes I is $\sigma_{I} = 100$ $\mu$Jy beam$^{-1}$, where the rms noise level of the polarized flux is $\sigma_{PI} = 20$ $\mu$Jy beam$^{-1}$. The difference in rms noise levels is due to the lower limit in the dynamic-range of the total intensity when compared to the polarized intensity. Then, the degree and position angle of polarization were estimated as $P = \sqrt{Q^{2} + U^{2}}$, $PA = 0.5\arctan{(U/Q)}$. The degree of polarization was corrected by bias. ALMA provides a systematic uncertainty in linear polarization of 0.03\%, which corresponds to a minimum detectible polarization of 0.1\%.
 

\section{Results}\label{sec:results}
 
\begin{figure*}[ht!]
\includegraphics[angle=0,scale=0.46]{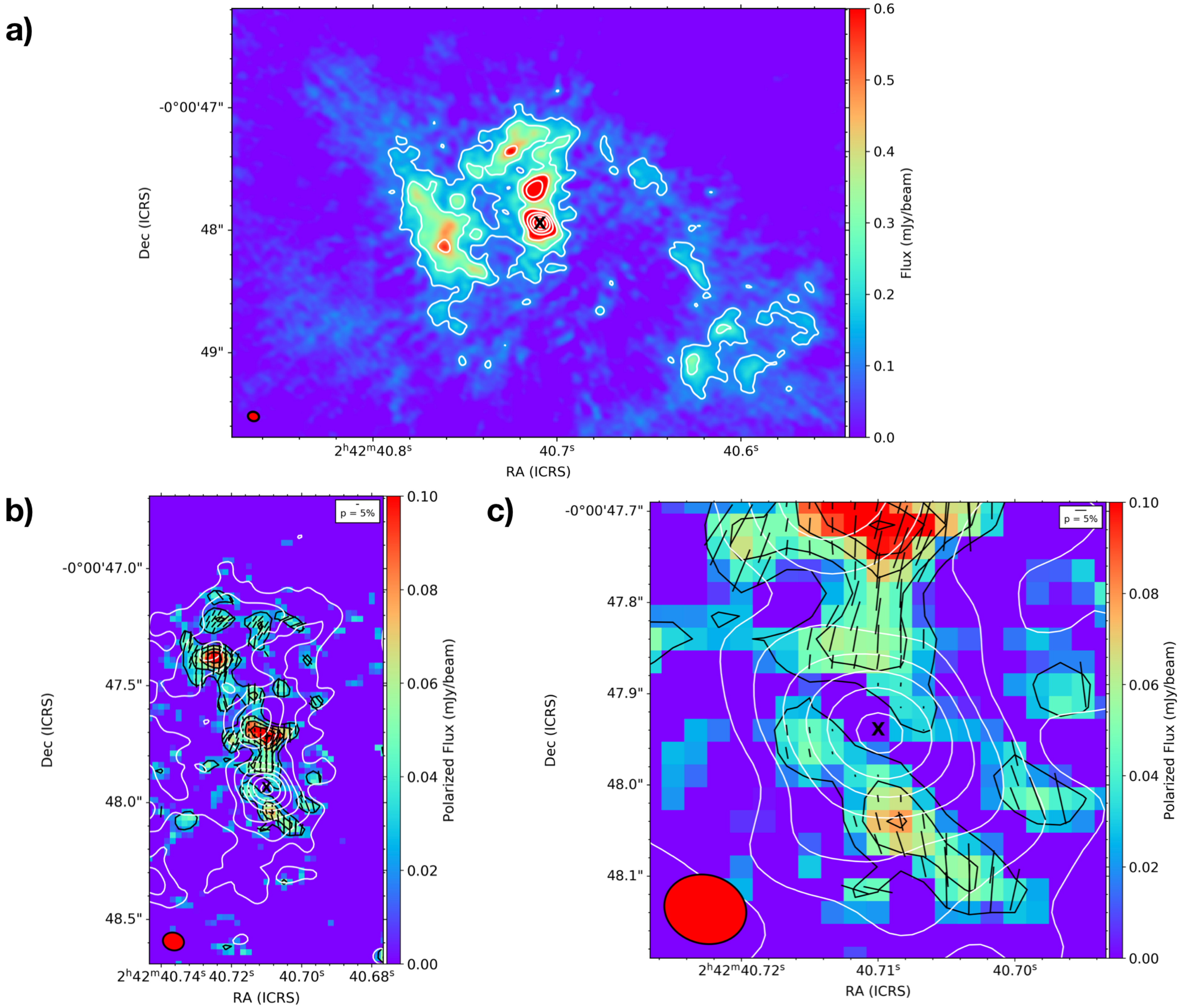}
\caption{\textit{a)} Total flux (color scale) image at 348.5 GHz (860 \um) of the central $5 \times 3.5$ arcsec$^2$ ($300 \times 210$ pc$^2$) region. Total flux contours (white) start at $4\sigma$ with $\sigma = 3.0 \times 10^{-5}$ Jy beam$^{-1}$ and increase as $2^{n}\sigma$, where $n = 2, 3, 4, \dots$
\textit{b)} Polarized flux (color scale) with overlaid polarization E-vectors (black) of the central $1 \times 1.5$ arcsec$^2$ ($60 \times 90$ pc$^2$) region. Total flux (white contours) as a). Polarized flux contours start at $3\sigma$ with $\sigma = 8.0 \times 10^{-3}$ Jy beam$^{-1}$ and increase in steps of $1\sigma$.
\textit{c)} Same as b) but within the central $0.5 \times 0.5$ arcsec$^2$ ($30 \times 30$ pc$^2$) region of NGC 1068. The beam (red) is shown in the lower-left of each figure, and a legend of a 5\% polarization vector is shown in the upper-right.
For all figures, the black cross shows the position of the AGN. }
\label{fig:fig1}
\epsscale{2.}
\end{figure*}
 
 Figure \ref{fig:fig1}-a shows the total flux image (continuum map) of NGC 1068 at 348.5 GHz (860 \um) within the central $5 \times 3.5$ arcsec$^2$ ($300 \times 210$ pc$^2$). Figure \ref{fig:fig1}-b shows the polarized flux with overlaid polarization E-vectors in a $1 \times 1.5$ arcsec$^2$ ($60 \times 90$ pc$^2$) and total flux contours as in Fig. \ref{fig:fig1}-a. Figure \ref{fig:fig1}-c shows the central $0.5 \times 0.5$ arcsec$^2$ ($30 \times 30$ pc$^2$) region. Each polarization vector shows the E-vector with a statistically independent measurement of $P/\sigma_{P} \ge 3$. The length of the vectors displays the polarization degree with a reference 5\% polarization vector shown in each figure. For each figure, the position of the AGN is indicated with a black cross.
 
 In the total flux image, we find an unresolved nucleus with additional extended emission along the N-S direction co-spatial with the jet direction, as well as along the East and South-West co-spatial with the circumnuclear disk (CND). These structures are consistent with previous findings by \citet{Imanishi2016, Imanishi2018} using ALMA Band 6. Here we present the results of the continuum and line emission of the torus, $<0\farcs2$, and the extended,  $>0\farcs2$, polarized emission.

 \subsection{The Total and Polarized Flux of the Torus}\label{subsec:cont}
 
\begin{figure*}[ht!]
\includegraphics[angle=0,scale=0.50]{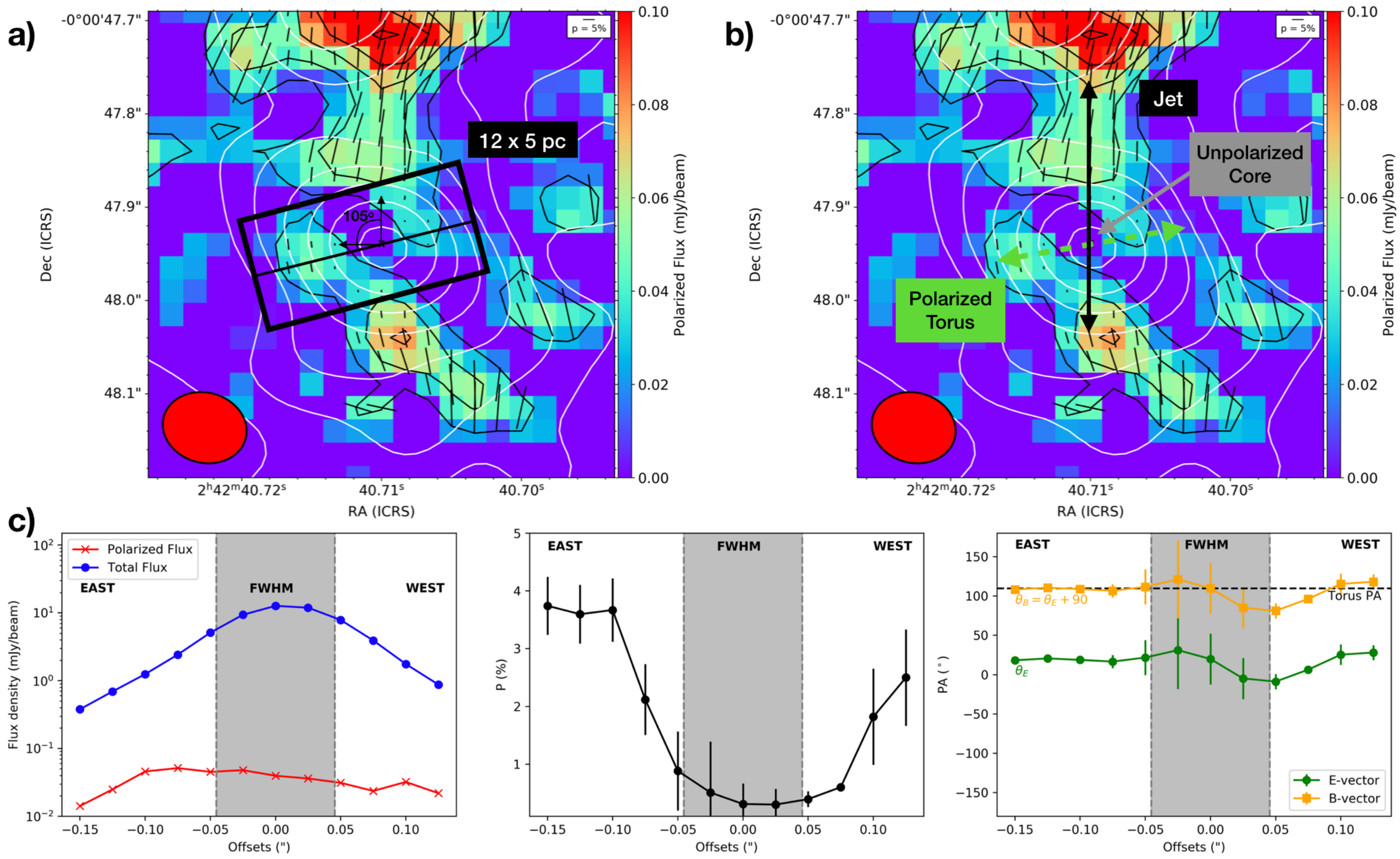}
\caption{\textit{a)} Same as Fig. \ref{fig:fig1}-c with the rectangular box showing the region used for the polarimetric analysis of the nucleus. 
\textit{b)} Physical interpretation of the polarization mechanisms in the central $0\farcs5 \times 0\farcs5$ ($30 \times 30$ parsec$^{2}$) of NGC 1068.
\textit{c)} Profiles of the torus along the long axis at a PA of $105^{\circ}$. \textit{Left:} Total (blue circles) and polarized (red crosses) fluxes. \textit{Middle:} Degree of polarization (black circles). \textit{Right:} PA of polarization of the E-vector (green circles) and B-vectors (E-vectors $+ 90^{\circ}$; orange squares). The PA $\sim 105^{\circ}$ of the long axis of the torus is shown (black dashed line). For all figures, the FWHM (grey shaded area) of the observations as well as the east and west sides of the torus are shown.
}
\label{fig:fig2}
\epsscale{2.}
\end{figure*}

The unresolved (0\farcs07, 5 pc) core is unpolarized at ALMA Band 7 (348.5 GHz, 860 \um) as shown in Fig. \ref{fig:fig1}-c. However, we have found a trace of a polarized component at $\sim0\farcs13$ (8 pc) along a PA of $\sim105^{\circ}$ East of North. To extract the total flux and polarization signature of the torus, we performed measurements of the total and polarized flux, and degree and PA of polarization within a $12 \times 5$ pc along a PA of $105^{\circ}$ region centered at the core of NGC 1068 (Fig. \ref{fig:fig2}-a). 

This region minimizes any potential contamination arising from the polarization of the jet in the northern and southern regions of the core, which maximizes the polarization signature of the torus (Figure \ref{fig:fig2}-a). The dimensions of this region is comparable with the size of the torus found by previous ALMA observations \citep[i.e.][]{GB2016,G2016,Imanishi2018,GB2019}. Specifically, we co-added the Stokes IQU at each location along the equatorial axis within bins of $25$ mas $\times$ $83$ mas ($1.5 \times 4.8$ pc). Then, the degree of polarization and polarized flux were estimated and debiased. Fig. \ref{fig:fig2}-c) shows the measurements of the total flux, polarized flux, and degree and PA of polarization along a PA of $105^{\circ}$ (i.e. extension of the torus). 

We measure that the polarized flux and the degree of polarization at the core (within the central beam) are consistent with null polarization. We find that the polarization increases as a function of the distance from the core along the equatorial plane (Fig. \ref{fig:fig2}-c). Specifically, we measure a statistically significant polarized peak of $3.7\pm0.5$\% in three consecutive bins from $0\farcs10$ ($6$ pc) to $0\farcs18$ ($10.8$ pc) along a PA of $105^{\circ}$ East from the AGN. To obtain an independent polarization measurement of the East region of the torus, we  estimate the polarization within a $0\farcs07 \times 0\farcs06$ ($4.2 \times 3.6$ pc) aperture centered at $0\farcs12$ and PA $= 105^{\circ}$ East from the AGN. The degree and PA of polarization were measured to be $2.2\pm0.3$\% and $19\pm2^{\circ}$, respectively. The jet, which has a different PA of polarization, shifts the PA and reduces the degree of polarization of the large aperture measurement when compared to the single bin measurement. The measured PA of polarization of the E-vector is nearly perpendicular to the equatorial plane of the torus, i.e. $\Delta PA = | (105\pm5)^{\circ} - (19\pm2)^{\circ}|= 86\pm6^{\circ}$. This result shows that the measured B-vector is parallel to the equatorial plane of the torus (Fig. \ref{fig:fig2}-c), i.e. the torus of NGC 1068 has a large-scale toroidal B-field.

 At Band 7, the non-thermal total flux emission is estimated to contribute $\sim30-65$\% within a 1\arcsec-aperture \citep{GB2014}. In polarized flux, \citet{GBO2004} found an unpolarized core at 5GHz for the non-thermal contribution. Given the high-angular resolution of our observations, we are able to disentangle the torus and jet in polarized emission, while we estimate that the total flux of the jet can contribute as much as $15$\% within a $0\farcs1$.

\subsection{The CN Emission Line of the Torus}\label{subsec:lines}

\begin{figure*}[ht!]
\includegraphics[angle=0,scale=0.43]{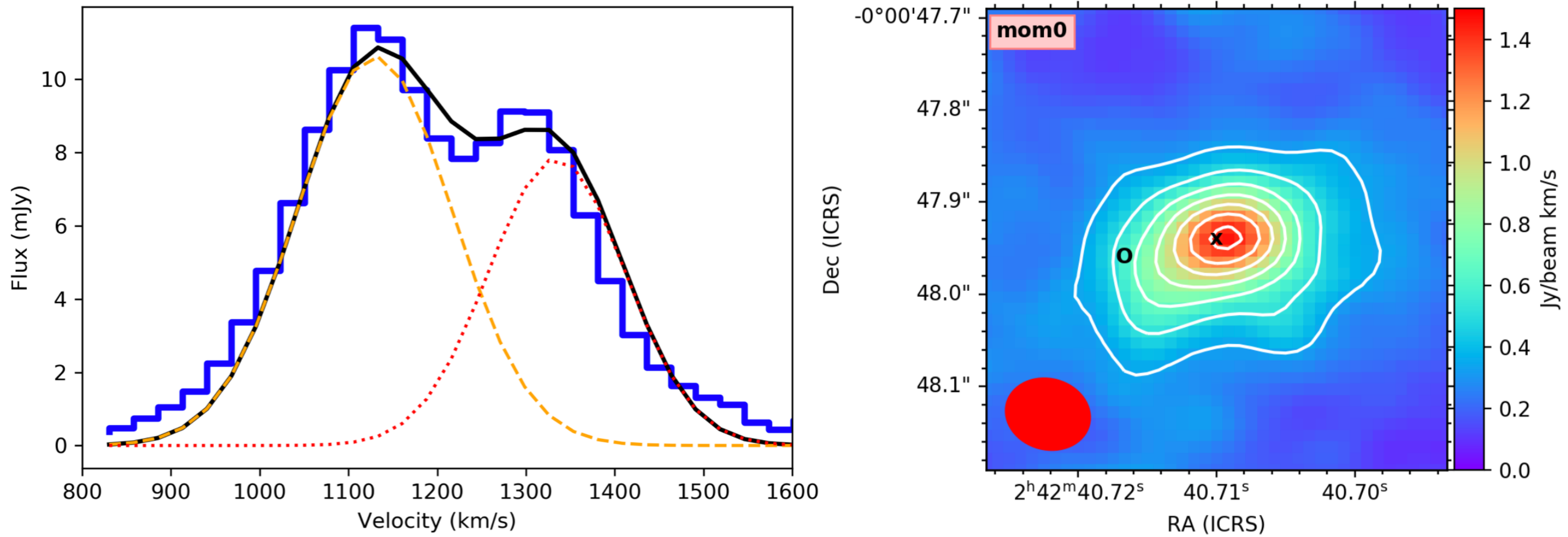}
\caption{\textit{Left:} Flux profile for the double line CN N=3-2 across the central $0\farcs5 \times 0\farcs5$ ($30 \times 30$ pc$^{2}$) region at a PA $= 110^{\circ}$. Best fit model to the flux profile using two Gaussians.
\textit{Right:} Integrated intensity (moment 0) of NGC 1068 of the combined double CN N=3-2 lines. Contours start at $3\sigma$ and increase in steps of $1\sigma$, where $\sigma = 0.1703$ Jy beam$^{-1}$ km s$^{-1}$.
 The mass-accreting SMBH position is indicated by a black cross, the highly polarized eastern location of the polarized region is indicated by a circle, and the beam size of the observations as a red ellipse.}
\label{fig:fig3}
\epsscale{2.}
\end{figure*}

We identify an emission line in our observations. Figure \ref{fig:fig3} shows the emission line within the $0\farcs5 \times 0\farcs5$ ($30 \times 30$ pc$^{2}$) region identified as CN N=3-2. We resolve the double line fine structure identified as CN N=3-2 J=7/2-5/2 and CN N=3-2 J=5/2-3/2. The integrated intensity (moment 0) of both lines is also shown in this figure, with the location of the position of the SMBH (black cross) and the polarized region (black circle) shown in Section \ref{subsec:cont}. We fit the emission lines with two 1D Gaussian profiles. The J=7/2-5/2 component has a peak flux of $10.62\pm0.32$ mJy, full width at half maximum (FWHM) of $86\pm5.5$ km s$^{-1}$, and total flux of $84.13\pm5.98$ mJy centered at $1129\pm7.1$ km s$^{-1}$. The J=5/2-3/2 component has a peak flux of $7.84\pm0.52$ mJy, FWHM of $76.8\pm6.1$ km s$^{-1}$, and total flux of $54.85\pm5.50$ mJy centered at $1334.8\pm8.3$ km s$^{-1}$. We estimate that $\sim15$\% of the J=7/2-5/2 component contributes to the J=5/2-3/2 component, and $\sim7$\% of the J=5/2-3/2 component contributes to the J=7/2-5/2 component. A detailed analysis to obtain the velocity fields, i.e. moments 1 and 2, of each line is difficult due to the blend of the fine structure. This emission line is also detected in the CND at scales of hundreds of pc by \citet{nakajima2015}; however these authors were not able to resolve the fine structure of this line due to the lower resolution of their observations.
 
The integrated intensity profile is concentrated along the west-east direction with an elongation of $\sim0\farcs3 \times 0\farcs15$ ($18 \times 9$ pc) at a PA of $100-110^{\circ}$. The peak of the integrated intensity is slightly shifted to the West, and there is more elongation towards the eastern than the western part.  This result shows similar physical conditions than the HCN J=3-2 and HCO$^{+}$ J =3-2 lines \citep{Imanishi2018}. These lines show a highly inhomogeneous torus along the equatorial axis with low velocity dispersion in the eastern part of the torus, while large velocity dispersion and strong emission are found in the western region of the torus. \citet{GB2016} found that the less dense molecular structure is lopsided. We conclude that our observed polarization region at 8 pc East from the core is co-spatial with 1) a low velocity dispersion region, and 2) a less dense molecular structure in the outermost eastern region of the molecular torus.

\subsection{The Total and Polarized Flux of the Extended Emission}\label{sec:OBSextended}

\begin{figure*}[ht!]
\includegraphics[angle=0,scale=0.42]{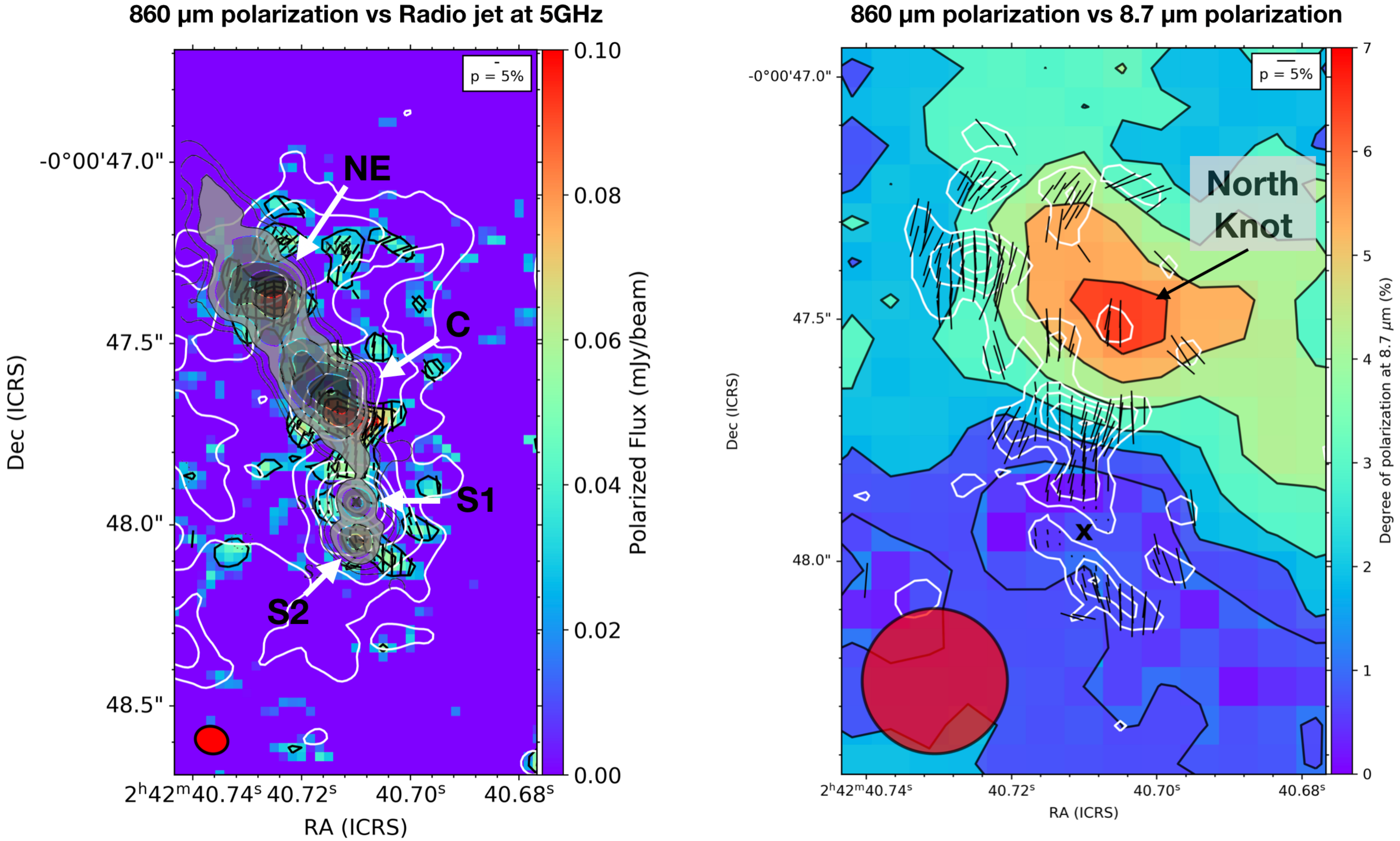}
\caption{\textit{Left:} 860 \um\ polarized flux (color scale) and polarization vectors (black vectors) with the overlaid radio jet emission (gray color scale) at 5 GHz \citep{GBO2004}. Total flux contours (white contours) as in Fig. \ref{fig:fig1} are shown. Each radio source is labeled and marked with a white arrow.
\textit{Right:} Degree of polarization (color contours) at 8.7 \um\ \citep{LR2016} with the overlaid 860 \um\ polarized flux (white contours) and polarization vectors  (black vectors). Degree of polarization contours in steps of 1\% are shown. The North Knot shows the jet-molecular cloud interaction. The mass-accreting SMBH position is indicated by a black cross. The beam size of $0\farcs38$ at 8.7 \um\ of CanariCam is shown (red circle).
For both figures, a legend of a 5\% polarization vector is shown.
}
\label{fig:fig4}
\epsscale{2.}
\end{figure*}

In polarized flux, we find resolved polarized emission structures along the N-S direction (Fig. \ref{fig:fig1}-b) within the central $1\times1.5$ sqarcsec ($60\times90$ pc$^{2}$). We measure polarization levels from $\sim2$\% to $\sim11$\% with the E-vectors of the PA of polarization to be mostly in the N-S direction. We overlaid (Fig. \ref{fig:fig4}) the 860 \um\ polarized flux and polarization vectors with the radio jet at 5 GHz observed with MERLIN \citep{GBO2004}. We find that the 860 \um\ polarized extended emission regions are spatially anti-correlated with the location of the radio knots, i.e. NE, C, and S2, of the radio jet at 5 GHz. 

We also overlaid the 860 \um\ polarized flux and polarization vectors with the polarization degree at 8.7 \um\ observed with CanariCam on the 10.4-m Gran Telescopio Canarias \citep{LR2016}. The polarization degree and polarized flux at 8.7 \um\ show a similar morphology \citep[fig. 2 and 3 of][]{LR2016}. We find that the 860 \um\ polarized flux is located at the boundaries of the eastern region of the MIR polarized extended emission identified as the `North Knot'. The North Knot is the region at $\sim0\farcs5$ ($30$ pc) North from the core of NGC 1068, where the jet changes direction from N-S to $\sim45^{\circ}$ East of North due to the interaction with a giant molecular cloud. At the center of the North Knot, we find a highly polarized, $\sim7$\%, region, which is at a similar level as the measured polarization, $\sim6.5$\%, at 8.7 \um. However, the PA of polarization at 860 \um\ differs from the PA of polarization at 8.7 \um, i.e. $\Delta PA= | PA_{860 \mu m} - PA_{8.7 \mu m}| = | 175\pm7^{\circ} - 44\pm3^{\circ}| = 131\pm8^{\circ}$. Section \ref{dis:radiojet} discusses the origin of the sub-mm polarization of these regions.


\section{Magnetic field model}\label{sec:Bmodel}

As mentioned in the introduction, the $2-13$ \um\ polarimetric observations of NGC 1068 suggest that the polarization arises from magnetically aligned dust grains from the dusty torus. However, effects of extinction by the torus itself produce a null net polarization due to self-absorption dichroic emission in the MIR \citep{LR2016}. At sub-mm wavelengths most of the dust is optically thin, which allows us to observe the bulk of emission of the dust grains in the torus.   We here develop a magnetic field model to study the measured polarization region of the torus and put constraints on the torus and magnetic field morphology.

\subsection{Model definition}\label{sec:modeldefinition}

A lot of effort has been put into successfully applied approaches \citep[e.g.][]{LD1985,FP2000,planckXX2015,CKL2016,king2018} that have computed the synthetic observations of the polarization emission in magneto-hydrodynamical (MHD) simulations. We here follow these standard approaches to estimate the expected polarization of the torus at $860$ \um\ by dust emission of magnetically aligned dust grains. We can express the Stokes parameters in terms of the local magnetic field $\textbf{B} = (B_{x}, B_{y}, B_{z})$ for optically thin, $\tau_{\lambda} << 1$, dust by the form

\begin{eqnarray}
I &=& \int{n \bigg( 1-p_{0} \bigg(  \frac{B_{x}^{2} + B_{y}^{2}}{B^{2}} - \frac{2}{3} \bigg) \bigg) }ds \\
Q &=& p_{0} \int{n \bigg( \frac{B_{y}^{2} - B_{x}^{2}}{B^{2}} \bigg)}ds \\
U &=& p_{0} \int{n \bigg( \frac{2B_{x}B_{y}}{B^{2}} \bigg)}ds
\end{eqnarray}
\noindent 
\citep{king2018}, where $x$ and $y$ define the plane of the sky,  $z$ is parallel to our line of sight (LOS),  $N = \int{n}ds$ is the column density, and $n$ is the volume density. We here define the quantity, $N_{2}$, as:

\begin{equation}
N_{2} = \int{n \bigg( \frac{B_{x}^{2} + B_{y}^{2}}{B^{2}} - \frac{2}{3}  \bigg)}ds
\end{equation}
\noindent
, thus Stokes I can be defined as

\begin{equation}
I = N - p_{0}N_{2}
\end{equation}
\noindent
where the term $N_{2}$ is a corrective factor that accounts for the decrease of emission by dust grains at a given inclination with respect to the plane of the sky \citep{FP2000}. This parameter provides a quantification of the level of dust grain alignment. For all grains aligned in the line of sight $N_{2} = -\frac{2}{3}N$, while for all grains aligned in the plane of the sky $N_{2} = \frac{1}{3}N$. In general, $N_{2}$ represents a small correction to the Stokes I and is of similar order to the dimensionless factor $p_{0}$. Thus, Stokes I (Eq. 5) is approximately the column density of the object. $p_{0}$ relates the grain cross sections and grains alignment properties, and it can be approximated as the maximum polarization by dust emission at the given wavelength. In sub-mm, the polarization by dichroic emission is not greater than 10\%, thus we adopt a $p_{0} = 0.1$ \citep{FP2000}.

The polarization fraction, $p$, and the polarization angle in the plane of the sky, $\chi$, are given by

\begin{eqnarray}
p &=& \frac{\sqrt{Q^2 + U^2}}{I} \\
\chi &=& 0.5 \arctan \bigg(\frac{U}{Q} \bigg)
\end{eqnarray}

The overall magnetic field of the torus is currently unknown. We rely on magnetohydrodynamical (MHD) simulations to assume the overall structure of the magnetic field in the torus. Recent MHD simulations by \citet{DK2017} assume an initial poloidal field in the torus. This poloidal field quickly and efficiently generates a toroidal field to maintain a long-lived torus due to differential rotation. More complex magnetic field configurations can also be assumed \citep[e.g.][]{aitken2002}, but here we will minimize the number of variables by assuming a toroidal structure to estimate the general behavior of the polarization as a function of the radius and optical depth. The toroidal configuration also presents a straightforward interpretation in terms of emission and absorption across the torus. For the magnetic field configuration of the torus, we assume a toroidal magnetic field in cartesian coordinates by the form

\begin{eqnarray}
B_{x} &=& B_{0}(r) \cos(\phi) \\
B_{y} &=& B_{0}(r) \sin(\phi) \\
B_{z} &=& B_{0}(r)  \\
\end{eqnarray}
\noindent
where $B_{0}(r)$ is the magnetic field as a function of the radial distance to the center, and $\phi$ is the azimuthal angle. When this magnetic field configuration is viewed at an inclination, $i$, and tilt angle, $\theta$, projected on the plane of the sky, the magnetic field at the observer's frame is given by

\begin{eqnarray}
B_{x^{s}} &=& B_{x}\cos\theta + (B_{y}\cos i - B_{z}\sin i)\sin\theta \\
B_{y^{s}} &=&  -B_{x}\sin\theta + (B_{y}\cos i - B_{z}\sin i) \cos\theta \\
B_{z^{s}} &=& B_{y}\sin i + B_{z}\cos i
\end{eqnarray}
\noindent
where ($x^{s},y^{s},z^{s}$) are the major axis, minor axis, and the LOS, respectively. Face-on view corresponds to $i$ = 0 $^{\circ}$ and edge-on to $i$ = 90 $^{\circ}$. Tilt axis has a reference point, $\theta = 0^{\circ}$, along the north-south direction and positively increases east from north. We assume a magnetic field distribution with a constant strength as a function of the radial distance to the center, $B_{0}(r) = const.$ (Section \ref{sec:Bstrength}).

\subsection{Synthetic observations}\label{sec:syntobs}

\begin{figure*}[ht!]
\includegraphics[angle=0,scale=0.44]{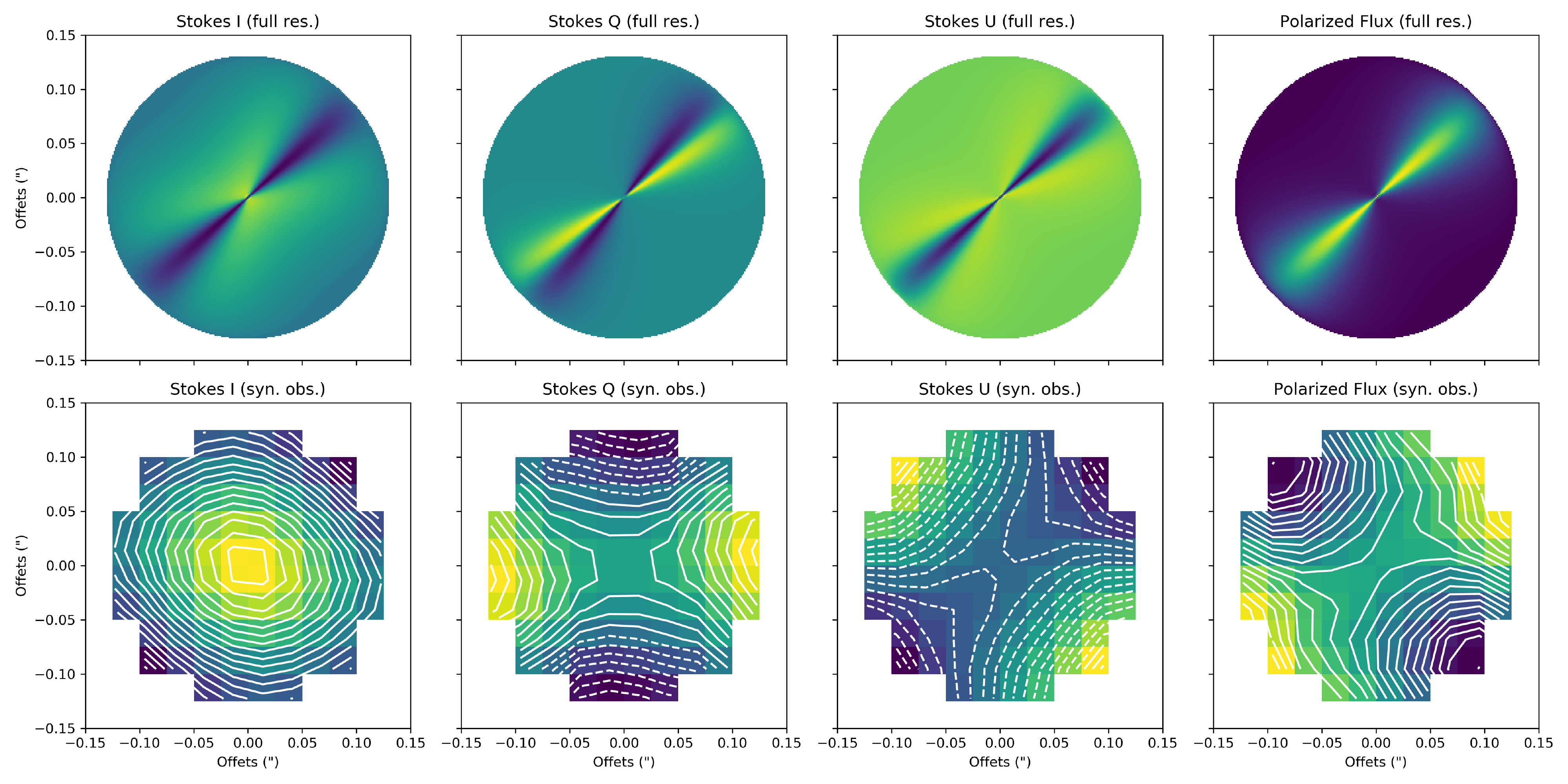}
\caption{Stokes I (first column), Q (second column), U (third column) and polarized flux (fourth column) for the polarization model of the torus at full resolution, $1.2$ mas (first row), and at the resolution and pixel size of $25$ mas of our ALMA observations of NGC 1068 (second row). 
}
\label{fig:fig5}
\epsscale{2.}
\end{figure*}

\begin{figure*}[ht!]
\includegraphics[angle=0,scale=0.40]{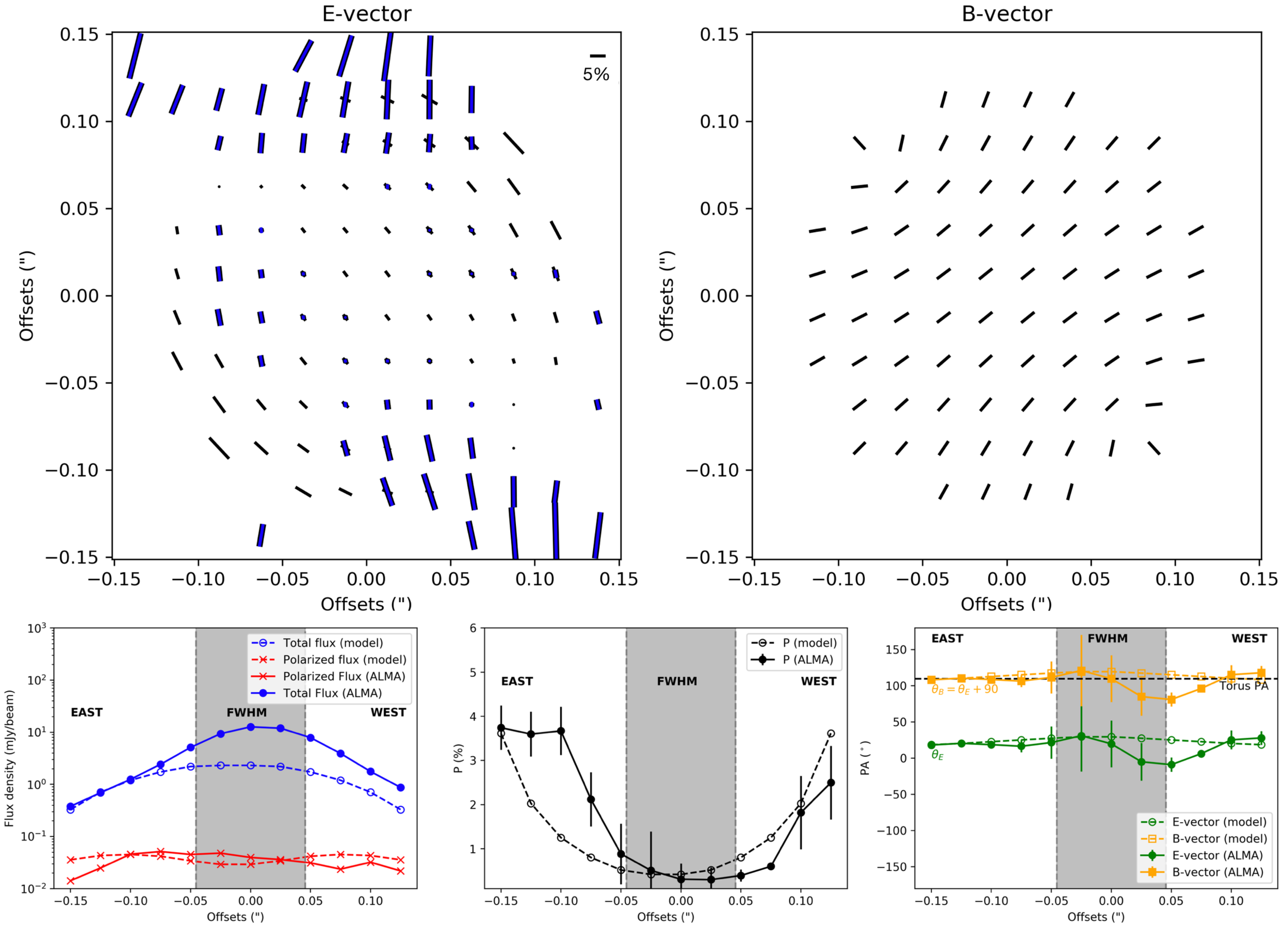}
\caption{\textit{Top}: Modeled E-vectors (left) and B-vectors (right) of the polarization model of the torus at the resolution of our ALMA observations. For comparison, the E-vectors measured with ALMA are shown (blue vectors). The length of E-vectors shows the degree of polarization and their orientation shows the PA of polarization. Length of B-vectors have been normalized and rotated by 90$^{\circ}$ to show the orientation of the magnetic field. 
\textit{Bottom:} Profiles of the observations (solid lines) and synthetic observations (dashed lines) along the long axis as shown in Fig. \ref{fig:fig2}.}
\label{fig:fig6}
\epsscale{2.}
\end{figure*}

To compute the column density of the torus per LOS, we use the surface brightness and cloud distribution computed by the radiative transfer code \textsc{clumpy} torus \citep{nenkova2008}. Specifically, we use the HyperCubes of AGN Tori (\textsc{hypercat}\footnote{\textsc{clumpy} torus images can be found at \url{https://www.clumpy.org/pages/images.html}}, R. Nikutta et al. in preparation). \textsc{hypercat} uses the \textsc{clumpy} torus models with any combination of torus model parameters to generate physically scaled and flux-calibrated 2D images of the dust emission and distribution for a given AGN. We use a distance of $14.4$ Mpc, and a torus tilt angle on the plane of the sky of $\theta = 110^{\circ}$ East of North based on the molecular torus morphology of our integrated flux of the CN line, the total flux continuum at Band 6 by \citet{GB2016}, and the molecular torus morphology observed at HCN J=3-2, HCO$^{+}$ = 3-2 by \citet{Imanishi2018}. To compute the 2D images of the dust continuum emission, we took the inferred torus parameters using the SED of NGC 1068 by \citet{LR2018}: angular width, $\sigma = 43^{\circ}$, radial thickness, $Y = 18$, number of clouds along the equatorial plane, $N_{0} = 4$, index of the radial density profile, $q = 0.08$, optical depth of each cloud at V-band, $\tau_{V} = 70$, and inclination angle, $i = 75^{\circ}$. These authors estimated a torus diameter of $10.2\pm0.4$ pc ($\sim0.2\arcsec$) with a height of $3.5^{+1.0}_{-1.3}$ pc ($\sim0.06\arcsec$), which is compatible with the sizes of the molecular torus shown in Fig. \ref{fig:fig3} and by \citet{GB2016,Imanishi2018}. We use the cloud distribution, i.e. number of clouds per LOS, multiplied by the optical depth at $860$ \um, as a proxy of the column density, $N$. We convert the optical depth per cloud from V-band, $\tau_V$, to $860$ \um\ by using the synthetic extinction curve\footnote{Synthetic extinction curve can be found at \url{https://www.astro.princeton.edu/~draine/dust/dustmix.html}} for $R_{V} = 5.5$ computed by \citet{WD2001}.
 
Using equations 1-3, a toroidal magnetic field configuration inclined at $i=75^{\circ}$ and tilted at $\theta=110^{\circ}$, and the 2D images of the continuum dust emission using the inferred torus parameters of  \citet{LR2018}, we compute the Stokes IQU at a full resolution of $1.2$ mas  (Figure \ref{fig:fig5}). The polarized flux is estimated as the Stokes I multiplied by the degree of polarization, $p$. To obtain the synthetic observations at the spatial resolution and instrumental configuration of the ALMA observations, we use the Stokes IQU at full resolution and convolved with the observed beam, $77 \times 58$ mas ($4.6 \times 3.5$ pc) and PA $= 75.13^{\circ}$, and pixelated to 25 mas pixel$^{-1}$. We obtain the synthetic Stokes IQU observations using CASA v5.4\footnote{CASA v5.4 can be found at \url{https://casa.nrao.edu/casadocs/casa-5.4.0}}, which computes the synthetic beam for the antenna configuration at the day of observations, takes into account the total integration time, and convolves it with the Stokes IQU at full resolution. Then, the final images are pixelated to the pixel scale, 25 mas pixel$^{-1}$, of the ALMA observations to obtain the synthetic Stokes IQU (Figure \ref{fig:fig5}). Finally, we compute the degree and PA of polarization and polarized flux. The polarization vectors (E-vectors) and magnetic field pattern (B-vectors) inferred from the synthetic Stokes IQU images are shown in Figure \ref{fig:fig6}. The length of E-vectors shows the degree of polarization, and their orientation shows the PA of polarization. In polarized dust emission, the measured E-vector is perpendicular to the direction of the magnetic field. We derive the B-vector map by rotating the E-vectors by $90^{\circ}$. Following the approach of our ALMA observations (Section \ref{subsec:cont}), we compute the profiles of the Stokes I, degree and polarized flux as a function of the radius at a PA of $105^{\circ}$ using the synthetic observations (Figure \ref{fig:fig6}). 

\subsection{Results of the magnetic field model}\label{sec:ModelResults}

We find that a toroidal magnetic field with an inclination of $75^{\circ}$, tilted at $110^{\circ}$ EofN, and the cloud distribution of the torus inferred by the nuclear SED \citep{LR2018} are able to explain the general polarization behavior of the 860 \um\ polarimetric observations across the equatorial plane of the torus within the central $0\farcs3 \times 0\farcs3$ ($18 \times 18$ pc) of NGC 1068. By default our polarimetric model is axisymmetric, thus the measured asymmetry across the equatorial plane is not explained by our model. This asymmetry is explained due to variations of a turbulent magnetic field in Section \ref{sec:Turbulence}. Fig. \ref{fig:fig6} shows that the measured trend of the polarization across the equatorial plane of the torus can be explained by magnetically aligned dust grains arising from an optically thin layer at the outer edge of the torus. 

In full resolution, Stokes I, a proxy of the column density $N$, shows that the equatorial plane and the core contains a thin layer of optically thick dust at 860 \um, where only continuum dust emission is observed in the outer parts of the torus. The synthetic observations show an unresolved torus with slightly larger FWHM than the beam of the observations due to the contribution of emission in the equatorial direction. The east-west extended emission is clearly observed in the polarized flux. The core is unpolarized, while the outermost part of the torus along the equatorial plane shows measurable polarized flux. The total flux profile (blue open circles in Fig. \ref{fig:fig6}) is flatter than the observed (blue closed circles) (Fig. \ref{fig:fig2}-b). We attribute this difference to the contribution of the sub-pc scale jet in the N-S direction. This component is not assumed in our model where both 1) the excess of total flux emission, and 2) the drop of polarization within the beam can be due to the mixed contribution of the polarization from the torus and from the jet. For 2), \citet{GBO2004} measured an unpolarized core from the radio jet at 5GHz, and we estimated a total flux contribution $<15$\% in the core of NGC 1068 at 860 \um. Although a multi-wavelength analysis using higher angular resolution observations can potentially disentangle both the jet and torus components within the $77 \times 58$ mas ($4.6 \times 3.5$ pc), we estimate that the measured polarized flux at $\sim8$ pc east from the core at PA $\sim105^{\circ}$ is dominated by dust emission of the torus.

\subsection{Exploration of torus model parameters}\label{sec:Modelexplore}

In this section we aim to put further constraints on the torus model geometry ﻿by exploring the synthetic polarimetric observations of NGC1068﻿ generated with different parameters of the \textsc{clumpy} torus models of﻿ \citet{nenkova2008}. This study provides an independent approach from﻿ modeling of the nuclear IR SED and MIR ﻿spectroscopy using the same models \citep[e.g.][]{AH2011,ichikawa2015,LR2018}. Using a toroidal magnetic field in all cases, the nominal torus parameters used in Section \ref{sec:syntobs} were explored, and the  figures in Appendix \ref{app:modelexp} show the most characteristic synthetic observations of each parameter range. 

We find that the inclination, $i$, angular width, $\sigma$, and the index radial density profile, $q$, produce significant changes in the polarization pattern of the synthetic observations. The optical depth of each cloud, $\tau_V$, and the number of clouds, $N$, have no implications on the synthetic observations because the torus is mostly optically thin at 860 \um. 

For the inclination, $i$, the toroidal magnetic field was also inclined for each model as shown in Eq. 12-14. We find that for a torus and a toroidal magnetic field at an inclination $i>65^{\circ}$, the observed polarization is reproduced with an uncertainty in the PA of polarization $<10^{\circ}$. At lower inclinations, the E-vectors deviate by $\sim90^{\circ}$ from the observed PA of polarization (Fig. \ref{fig:fig7},\ref{fig:fig8}). The degree of polarization for low inclinations, i.e. almost face-on views, produces a double dip in the degree of polarization due to the difference of polarization angles as a function of the azimuthal angle. The torus inclination must be close to edge-on, $i \sim 90^{\circ}$, to reproduce the observations.

For angular widths $\sigma < 30^{\circ}$ (Fig. \ref{fig:fig9}), high ($>5$\%) polarization is measured across the  torus  and the PA of polarization differs by $>10^{\circ}$ from our measurements. Given our $7\sigma$ polarization detection, the uncertainty in our measured PA of polarization is $\pm4^{\circ}$. The height of the torus is compact and most of the dust is in a thin layer across the equatorial plane. Our observations have enough sensitivity to resolve this configuration, however a torus with $\sigma < 30^{\circ}$ does not reproduce the observations. For angular widths $\sigma > 60^{\circ}$ (Fig. \ref{fig:fig10}), the height of the torus is wide, which produces a low ($<1$\%) polarization across the torus. Although this configuration produces a comparable degree of polarization, the PA of polarization at the core differs $>10^{\circ}$ from our measurements, which does not reproduce the observations. Based on the measured polarization, we find that the torus angular width must be in the range of $33^{\circ}-50^{\circ}$ to reproduce the observations.

For index radial profiles $q > 1$, the torus became very compact, with most of the dust concentrated within the central few pc (Fig. \ref{fig:fig11}). This configuration produces a highly polarized and unresolved core, which does not reproduce the observations. The torus index radial profiles must be $q << 1$ to reproduce the observations.

For radial thickness $Y < 10$, the torus size will be $<6$ pc ($0\farcs1$), which will be unresolved by our observations as well as incompatible with previous ALMA observations \citep[i.e.][]{GB2016,Imanishi2018}. Our observations implied a torus with a radial thickness of $Y\ge17$, which implies a torus radius $\ge 9$ pc.

To sum up, we find that to reproduce the 860 $\mu$m polarimetric observations, the torus 1) is highly inclined, $i>65^{\circ}$, 2) has intermediate angular widths $\sigma = 33-50^{\circ}$, 3) has a flatter radial distribution of clouds, $q<<1$, and 4) has a radius $\ge9$ pc ($Y\ge17$).


\section{Discussion}\label{sec:DIS}

\subsection{A torus with magnetically aligned dust grains}

We conclude that the measured 860 \um\ polarization arises from magnetically aligned  dust grains in the torus. Our magnetic field model suggests that a constant toroidal magnetic field distribution inclined at $75^{\circ}$ with a size up to $18$ pc diameter produces the alignment of the elongated dust grains, and thus the measured polarization. This configuration produces a coherent PA of polarization (E-vectors) along the equatorial plane, which is consistent with our measured PA of polarization in the eastern part of the torus. Our polarization model suggests that the degree of polarization varies as a function of the optical depth, a known behavior of the polarization emission in molecular clouds from Far-IR (FIR) to sub-mm \citep{H1999,HK2004}. The polarization reaches a minimum at the location of the core where the AGN is highly extinguished, i.e. high optical depths. However, the degree of polarization increases at large radii and heights, i.e. low optical depths. The E-vector polarization map (Fig. \ref{fig:fig6}) shows this behavior. Although the expected theoretical polarization at the core is above the minimum detectable polarization by ALMA at Band 7, the effects of 1) optical depth, 2) angular resolution, and 3) contamination of the sub-pc jet make the core unpolarized. Only the outermost regions of the torus are less influenced by these effects because the torus is spatially resolved and  optically thin.

\subsection{A turbulent torus}\label{sec:Turbulence}

The magnetic field has both a constant and turbulent component \citep[e.g.][]{planck2015a}. The effects of the turbulent component can be studied through the variations of the PA of polarization using a structure function \citep{H2009}, as well as trends in the degree of polarization with the optical depth \citep{H1999}. Our observations only detect a significant polarization in the eastern part of the torus, which is not fully explained by only using our magnetic field model. 

As mentioned in the introduction and Section \ref{subsec:lines}, CN J=5-2, HCN J=3-2 and HCO$^{+}$ J =3-2 lines indicate that the torus is highly inhomogeneous along the equatorial axis. Specifically, low velocity dispersion is found in the eastern part of the torus, while large velocity dispersion and strong emission are found in the western region of the torus \citep{Imanishi2018}. Both observations and theoretical models \citep{MSI2007} suggest that CN is found in abundance at early stages of molecular cloud evolution, or when the region is irradiated by intense UV and X-ray photons. Under these conditions, CN can be found in several scenarios: 1) in high-temperature conditions, the reaction CN + H$_{2}$ $\rightarrow$ HCN + H efficiently converts CN into HCN \citep{HHW2010}, and 2) in regions with high turbulence can efficiently mix ionized/atomic gas and increase the amount of CN.  In general, any perturbance in the velocity fields introduces turbulence in the magnetic field. The detection of CN by our observations and the fact that the CN is associated with regions with high turbulence, high density clouds, and intense UV and X-ray radiation, may explain the behavior of the degree of polarization across the equatorial plane. Specifically, the measured gradient of the velocity dispersion along the equatorial plane makes the turbulent component of the magnetic field dominate in those areas with high velocity dispersion. For regions where the velocity dispersion is low, the coherent magnetic field may dominate, which can enhance the grain alignment and produce a higher degree of polarization. We conclude that the asymmetry in the degree of polarization along the equatorial axis can be explained as an increase in the level of turbulence, i.e. increase of the magnetic field turbulence at sub-pc scales, from the eastern to the western side of the torus. 

\subsection{The magnetic field strength of the torus}\label{sec:Bstrength}

The torus of NGC 1068 has been observed to have a gradient in velocity dispersion along the equatorial plane that may produce a variation of the turbulent magnetic field. To infer the relative contribution of both components, \citet{H2009} show that the ratio of turbulent to large-scale  components of the magnetic field strength, $\langle B^{2}_{t}\rangle^{1/2} / B_{o}$, is a function of the dispersion of polarization angles on the plane of the sky, $\alpha$, given by

\begin{equation}
\frac{\langle B^{2}_{t}\rangle^{1/2}}{B_{0}} = \frac{\alpha}{\sqrt{2-\alpha^2}}
\end{equation} 

Based on our 860 \um\ polarimetric observations, we estimate a dispersion of the PA of polarization of $\alpha = 5 \pm 2^{\circ}$ ($0.0873\pm0.0349$ radians). We estimated the dispersion and its associated error using the variations of the measured PA of polarization per bin in the eastern side of the torus from 0\farcs05 (3 pc) to 0\farcs20 (12 pc) shown in Fig. \ref{fig:fig6}. The turbulent to constant ratio of the magnetic field strength is estimated to be $\langle B^{2}_{t}\rangle^{1/2} / B_{o} = 0.062\pm0.025$ within a $0\farcs05-0\farcs15$ ($3-8$ pc) region of the eastern part of the torus.

The Davis-Chandrasekhar-Fermi method \citep[DCF method hereafter, ][]{D1951,CF1953} provides an empirical estimation of the plane-of-the-sky magnetic field strength. This method relates the magnetic field strength, $B$, with the dispersion of the polarization angles, $\alpha$, of the constant component of the magnetic field, and the velocity dispersion of the gas, $\sigma_V$. This method assumes equipartition and a constant component of the magnetic field. A modified version of the DCF method takes into account the characteristic turbulent-to-ordered ratio, which allows us to estimate the strength of the large-scale magnetic field as

\begin{equation}
B_{o} \simeq \sqrt{8\pi\rho} \frac{\sigma_{v}}{\alpha}
\end{equation}
\noindent
\citep{H2009}, which is valid when $B_{t} << B_{o}$, and where $\rho$ is the volume mass density in g cm$^{-3}$, and $\sigma_{V}$ is the velocity dispersion in cm s$^{-1}$.

The velocity dispersion of the molecular gas at the location of the polarized region in our observations is estimated to be $\sigma_{v, HCN 3-2} = 40$ km s$^{-1}$ and $\sigma_{v, HCO^{+} 3-2} = 10$ km s$^{-1}$ using ALMA observations  \citep{Imanishi2018}. As mentioned in the previous section, the measured polarization spatially corresponds with regions of low velocity dispersion. We take a velocity dispersion of $\sigma_{V} = 10$ km s$^{-1}$. \citet{GB2016} estimated a dust mass of the torus to be $(1.6\pm0.2) \times 10^{3}$ M$_{\odot}$, which assumes that other mechanisms rather that thermal emission contribute up $\sim18$\%. \citet{LR2018} estimated a torus diameter of $10.2\pm0.4$ pc and a height of $3.5^{+1.0}_{-1.3}$ pc yield a median volume mass density of $(1.4^{+1.43}_{-0.62}) \times 10^{-22}$ g cm$^{-3}$. Based on our 860 \um\ polarimetric observations, we estimate a dispersion of the PA of polarization of $\alpha = 5 \pm 2^{\circ}$ ($0.0873\pm0.0349$ radians). We estimate a magnetic field strength of $B = 0.67_{-0.31}^{+0.94}$ mG within the $3-8$ pc region of the eastern side along the equatorial plane of the torus. Note that we have used the median volume mass density of the whole torus, while the location of the polarized region of the torus corresponds to the less dense areas of the torus. Thus, we estimate an upper-limit of the large-scale magnetic field strength in the eastern size of the torus.

Our estimate of $\langle B^{2}_{t}\rangle^{1/2} / B_{o}$ shows that within the eastern part of the torus, at $3-8$ pc from the core, there seems to be an ordered region with small turbulent components ($\sim0.06$ times $B_{o}$). The magnetic field strength at a distance of 0.4 pc of NGC 1068 was estimated to be $52^{+4}_{-8}$ mG \citep{LR2015} with the $\langle B^{2}_{t}\rangle^{1/2} / B_{o}$ estimated to be $0.1$.  To conserve the magnetic field flux, assuming perfect conductivity, the magnetic field strength varies as a function of the radial distance as $B \propto r^{-2}$. Taken the values estimated by \citet{LR2015}, we estimate a magnetic field strength at 8 pc of $\sim0.14$ mG, which is in agreement  within the uncertainties of our measurements. We note that a magnetic field distribution with a variable strength as a function of the radial distance to the center ($B_{0}(r)$ in Section \ref{sec:modeldefinition}) will be required to study future high-angular resolution polarimetric observations as well as MHD modeling of AGN. 

The results presented in this section indicate that at sub-pc scales the turbulent component of the magnetic field is higher than at pc-scales. The radiative pressure dominates at sub-pc scales, while at larger distances (few pc) the magnetic field is more coherent.  We conclude that a large-scale magnetic field seems to be present in the torus from 0.4 to 8 pc, with a decrease in the turbulence component as the radius increases along the equatorial plane of the torus. At pc-scale, the magnetic fields can dominate the dynamics of the torus.

\subsection{Alternative polarization mechanisms}\label{sec:DisAlt}

We explore other polarization mechanisms to explain the measured 860 \um\ polarization at the core of NGC 1068. The polarization signature of synchrotron polarization is dominated by a high degree of polarization, and a PA of polarization of the E-vector perpendicular to the magnetic field direction. Although the measured PA of polarization, $19\pm2^{\circ}$, is tentatively parallel to the north-south direction of the jet, the location of the measured polarization, $\sim8$ pc east from the core, and the measured degree of polarization, $2.2\pm0.3$\%, rule out this mechanism. Self-scattering by dust has been proposed as a dominant polarization mechanism in protoplanetary disks observed with ALMA \citep[e.g.][]{yang2016,stephens2017}. In Band 7 and for HL Tau (a protoplanetary disk inclined at $\sim45^{\circ}$), a degree of polarization with a dip and PA of polarization almost perpendicular to the long-axis of the disk are observed. These studies suggest that self-scattering by large dust grains of sizes $> 50$ \um~are responsible for the observed polarization. Most of the studies characterizing the dust grains sizes in AGN point to sub-\um\ to \um\ dust grain sizes \citep[i.e.][]{LSL2008,XLH2017,SJL2017}. Although the measured polarization and its pattern are compatible to our polarimetric observations, the requisite physical conditions of the dust grain sizes are not present, and we can rule out self-scattering mechanisms as the dominant polarization.

\subsection{The extended polarized emission}\label{dis:radiojet}

Although the goal of this paper is a detailed analysis of the polarization signature of the dusty torus, we find highly polarized, up to 11\%, emission in the extended regions of NGC 1068. A thorough analysis of this extended polarized emission is outside the scope of this paper, but here we identify the several polarization regions and provide an interpretation of the potential dominant polarization mechanisms.

We found an almost uniform PA of polarization along the N-S direction with degree of polarization in the range of 2\% to 11\% from $0\farcs1$ to $0\farcs3$ ($6-18$ pc) North from the core. \citet{GB2019} derived the power-law index ($\alpha$, $S_{\nu} \propto \nu^{\alpha}$) in the range of $-0.79$ to $-0.26$, which is consistent with the $\alpha = -0.67$ at $1-12$ GHz by \citet{G1996}. These estimations were performed within the  $344.5-694$ GHz continuum emission using ALMA observations over an area of $0\farcs11 \times 0\farcs11$ ($6.6$ pc$^{2}$) centered at knot C.  Using these power-law indexes, we estimate an intrinsic synchrotron polarization to be close to the theoretical $65$\% polarization. However the measured polarization from our observations range between $2-11$\%. This area is co-spatial with intense UV emission \citep[e.g.][]{K1999} and dust emission at MIR wavelengths \citep[e.g.][]{T2001,LR2016}. There may be selective extinction across this area that decreases the intrinsic polarization by synchrotron emission. The inferred magnetic field morphology of the synchrotron emission is perpendicular to the measured PA of polarization. A helical magnetic field can likely explain the morphology right above the core, while a more complex structure, i.e. bow shock, at the arc-shape area of this knot is required. Another alternative can be radiative dust grain alignment \citep[RAT,][]{HL2009} due to the intense UV emission from the AGN. Further modeling is required to test these hypothesis.

We measure a highly polarized, $\sim7$\%, region spatially coincident with the North Knot detected at $8.7$ \um. The North Knot polarization decreases from $\sim7$\% to $\sim4$\% in the $8.7-11.6$ \um\ wavelength range \citep{LR2016}, which was interpreted as polarization arising from dust and gas emission heated by the post-shock gas after the jet-molecular cloud interaction. The dust emission is then further extinguished by aligned dust grains in the inner bar at $\sim45^{\circ}$ EofN of the host galaxy. At sub-mm wavelengths, the dust in the inner bar is optically thin. Therefore, the measured 860 \um\ polarization of the North Knot is the intrinsic polarization by thermal emission from magnetically aligned dust grains.

We found several highly polarized, up to 11\%, regions at $\sim0\farcs7$ ($\sim42$ pc) NE from the core. The NE knot has a spectral index of $-1.04$ at $1-12$ GHz \citep{G1996}, and $-1.6$ at  694 GHz- 344.5 GHz \citep{GB2019}, which is consistent with synchrotron emission and comparable to the spectral index at the large-scale of the jet \citep{WU1987}. The measured 860 \um\ polarization is in the range of $5-11$\% with a PA of polarization almost N-S, with a change of the degree of polarization along the western part of the NE knot (where the peak polarization is located). Using a spectral index of $-1.6$, we estimated an optically thin synchrotron emission polarization of $\sim60$\%, and a $\sim9$\% for optically thick synchrotron emission. The difference in polarization under the optically thin regime and the comparable polarization of the optically thick synchrotron emission indicates a strong depolarization due to Faraday rotation on the NE knot. The difference of the degree and PA of polarization in the western part of the knot may indicate the presence of a shock by the jet with the ISM, and/or a disruption of the knot in the radio jet.


\section{Conclusions}\label{sec:CON}

We performed $0\farcs07$ ($4.2$ pc) resolution ALMA Cycle 4 polarimetric Band 7 ($348.5$ GHz, $860$ \um) observations of the nuclear region of NGC 1068. Thanks to the high-angular resolution we have spatially resolved the jet emission along the north-south direction and the torus emission along the east-west direction. We also reported the detection of CN N=3-2 line in the central 16 pc, which is morphologically coincident with the molecular torus found at Band 6, HCN 3-2 and HCO$^{+}$ 3-2 ALMA observations. We detected the polarization signature of the torus by means of magnetically aligned dust grains emission. The polarized torus has a variation of the degree of polarization as a function of the radius along the equatorial plane, with a peak polarization of $3.7\pm0.5$\% and PA of polarization of $19\pm2^{\circ}$ (E-vector) at $\sim8$ pc east from the core. The core, at the location of the SMBH and peak of the total intensity, is unpolarized. This result is attributed to a combination of the increase of column density in our LOS towards the core, and contamination of the sub-pc scale jet and torus polarized emission within the beam. 

We computed synthetic polarimetric observations of magnetically aligned dust grains assuming a toroidal magnetic field and homogenous dust grain alignment. This model allows us to estimate the expected degree and PA polarization by dust emission in the torus. We find that to reproduce the 860 \um\ polarimetric observations, the torus 1) is highly inclined, $i>65^{\circ}$, 2) has intermediate angular widths $\sigma = 33-50^{\circ}$, 3) has a relatively flat radial distribution of clouds, $q<<1$, and 4) has a radius $\ge9$ pc ($Y\ge17$). The torus and the magnetic field distribution are inclined at $75^{\circ}$ with a size up to $18$ pc diameter. These results are consistent with the torus model parameters inferred from the IR SED modeling and provide an independent approach to physically constrain the torus morphology.

The measured asymmetry in the degree of polarization as a function of the radius across the equatorial axis can be explained by gradient of velocity dispersion, ie. variation of the magnetic field turbulence, from the eastern to the western part of the torus. We computed a turbulent-to-ordered ratio of the magnetic field of $\langle B^{2}_{t}\rangle^{1/2} / B_{o} = 0.062\pm0.025$ within a $0\farcs05-0\farcs15$ ($3-8$ pc) region of the eastern part of the torus, with a magnetic field strength estimated to be $0.67^{+0.94}_{-0.31}$ mG. When compared with $2.2$ \um\ polarimetric observations, $52^{+4}_{-8}$ mG and $\langle B^{2}_{t}\rangle^{1/2} / B_{o} \sim 0.1$, a gradient in the turbulent magnetic field as a function of the radius to the core is found. We concluded that a large-scale magnetic field may be present from $0.4$ to $8$ pc with a decrease of the turbulent component as radius increases along the equatorial plane of the torus. 

Our observations show that the bulk of the  total and polarized dust emission in the torus at sub-mm wavelengths is located along the equatorial direction. This result is well explained by dusty tori models in conjunction with a large-scale magnetic field within the central 16 pc of NGC 1068. The analysis presented here can be used in a larger sample of AGN performed at (sub-)pc scales using multi-wavelength total and polarimetric observations. Further dust emission and MHD modeling are required to study the turbulent magnetic fields in the torus.


\acknowledgments


This paper makes use of the following ALMA data: ADS/JAO.ALMA\#2016.1.00176.S. ALMA is a partnership of ESO (representing its member states), NSF (USA) and NINS (Japan), together with NRC (Canada), MOST and ASIAA (Taiwan), and KASI (Republic of Korea), in cooperation with the Republic of Chile. The Joint ALMA Observatory is operated by ESO, AUI/NRAO and NAOJ. The National Radio Astronomy Observatory is a facility of the National Science Foundation operated under cooperative agreement by Associated Universities, Inc. This research was conducted at the SOFIA Science Center, which is operated by the Universities Space Research Association under contract NNA17BF53C with the National Aeronautics and Space Administration. AA-H and S.G.-B. acknowledge support through grant PGC2018-094671-B-I00 (MCI/AEI/FEDER,UE). AA-H work was done under project No. MDM-2017-0737 Unidad de Excelencia "Mar\'ia de  Maeztu"- Centro de Astrobiolog\'ia  (INTA-CSIC).

%

\vspace{5mm}
\facilities{ALMA Polarimetry (Band 7)}


\software{\textsc{astropy} \citep{2013A&A...558A..33A}
\textsc{CASA} \citep{mcmullin2007}
          }



\appendix

\section{Magnetic field model parameter exploration}\label{app:modelexp}

This section shows the synthetic polarimetric observations produced by the magnetic field model shown in Section 4. We use the same torus model parameters as in Section \ref{sec:Modelexplore}, where only one parameter was changed at the time for each figure.

\begin{figure*}[ht!]
\includegraphics[angle=0,scale=0.80]{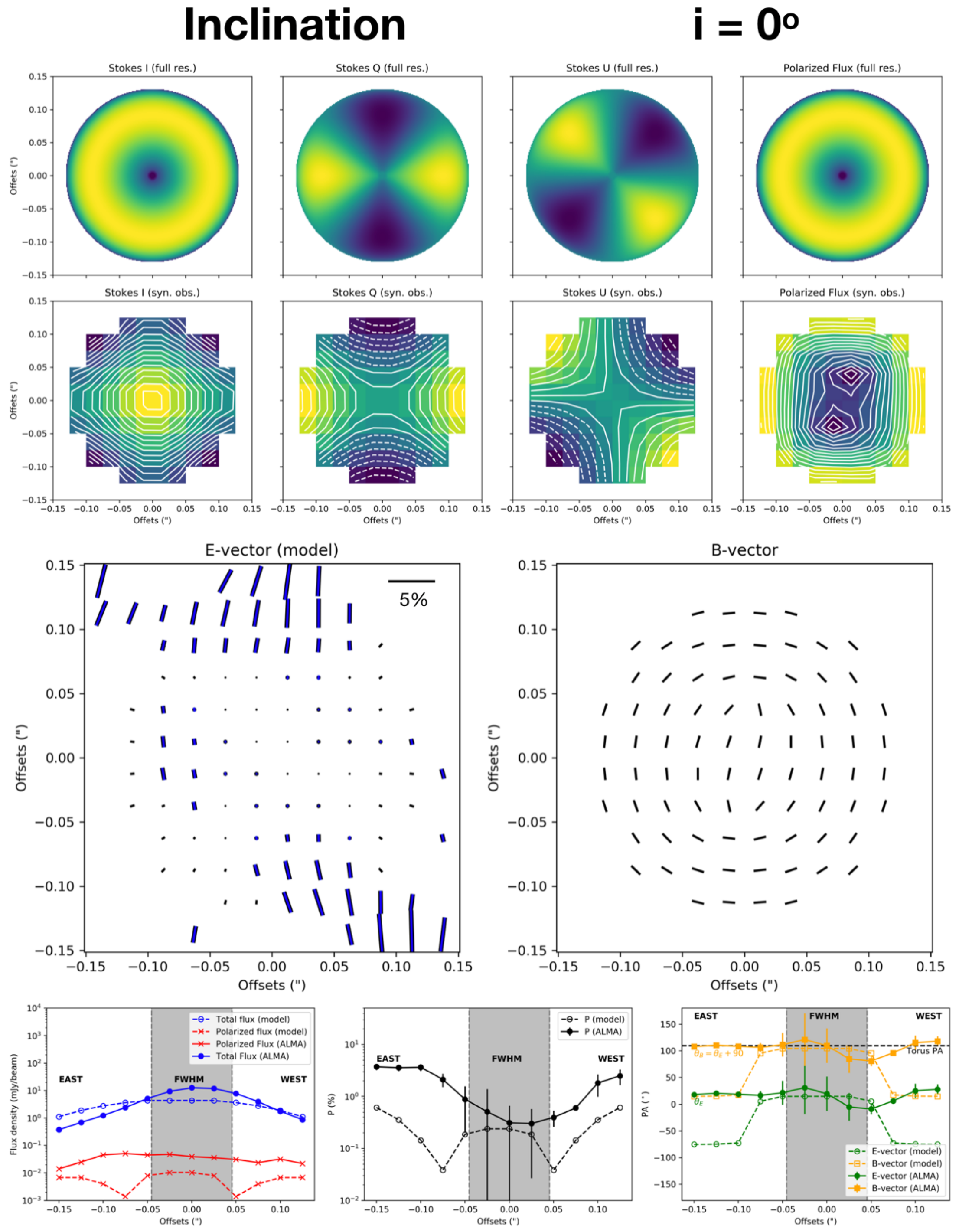}
\caption{Magnetic field model for an inclination of $i=0^{\circ}$. Full resolution and synthetic observations matching our ALMA observations of NGC 1068 are shown as in Fig. \ref{fig:fig5}. Polarization pattern of the E-vectors and B-vectors and their total, polarized flux, degree and PA of polarization as shown in Fig. \ref{fig:fig6}. At low ($<65^{\circ}$) inclination angles, the position angle of polarization of the E-vector does not reproduce the ALMA observations.
}
\label{fig:fig7}
\epsscale{2.}
\end{figure*}

\begin{figure*}[ht!]
\includegraphics[angle=0,scale=0.80]{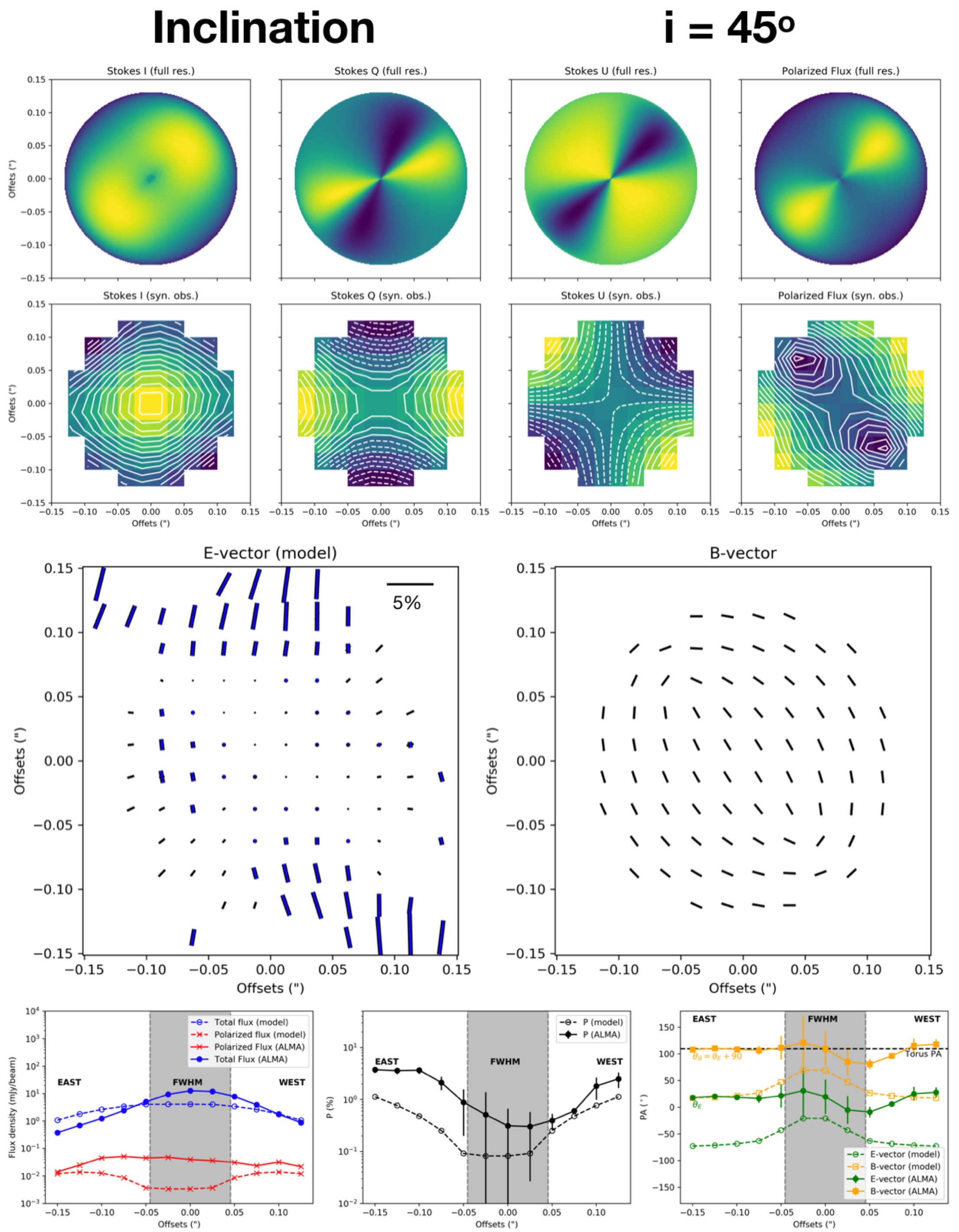}
\caption{Magnetic field model for an inclination of $i=45^{\circ}$. Same figures as Fig. \ref{fig:fig7}. At low ($<65^{\circ}$) inclination angles, the position angle of polarization of the E-vector does not reproduce the ALMA observations.
}
\label{fig:fig8}
\epsscale{2.}
\end{figure*}

\begin{figure*}[ht!]
\includegraphics[angle=0,scale=0.80]{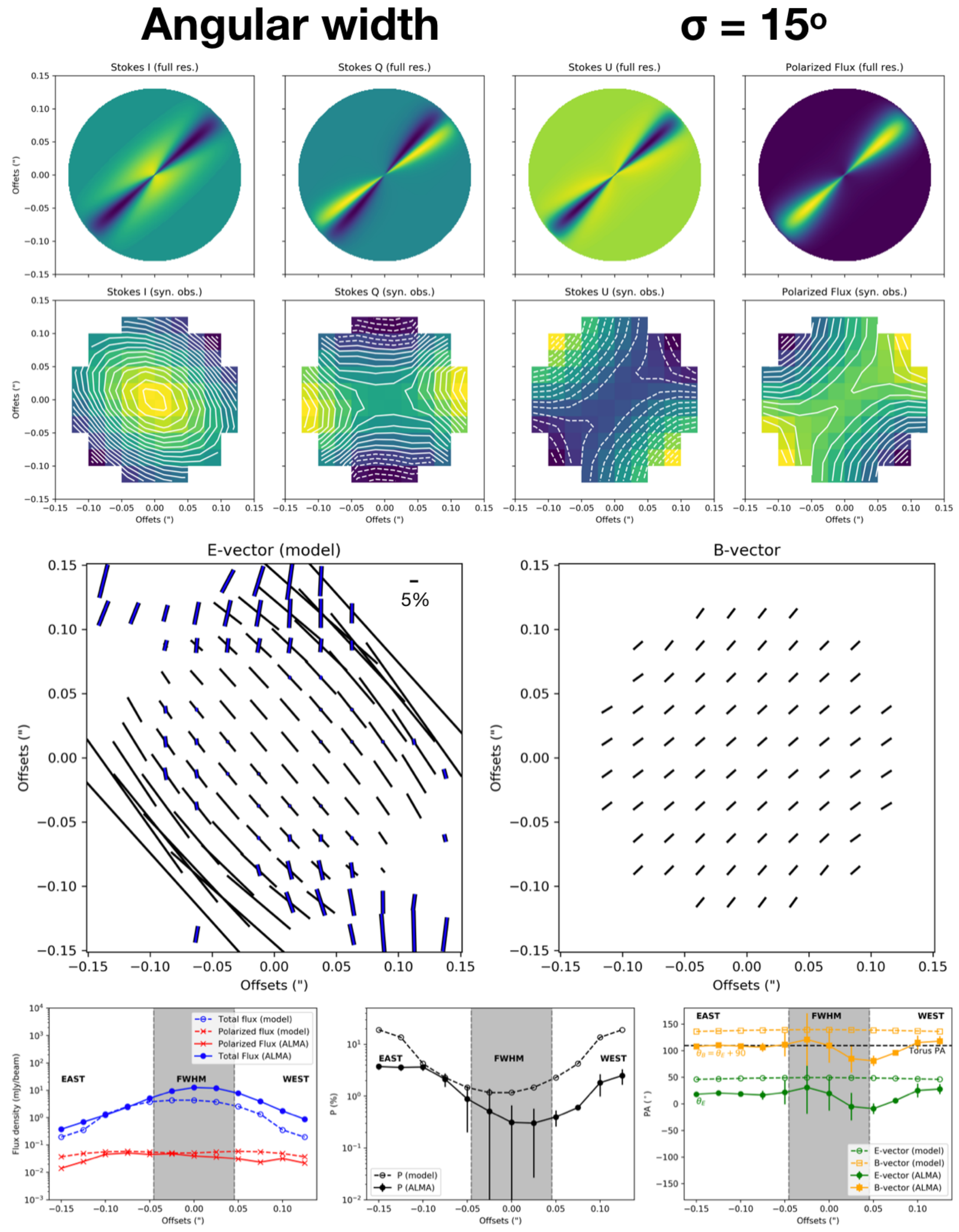}
\caption{Magnetic field model for an angular width of $\sigma = 15^{\circ}$. Same figures as Fig. \ref{fig:fig7}. For $\sigma < 30^{\circ}$, polarization is only observed at the outer edge. The polarization is higher than what has been observed in this work. 
}
\label{fig:fig9}
\epsscale{2.}
\end{figure*}

\begin{figure*}[ht!]
\includegraphics[angle=0,scale=0.80]{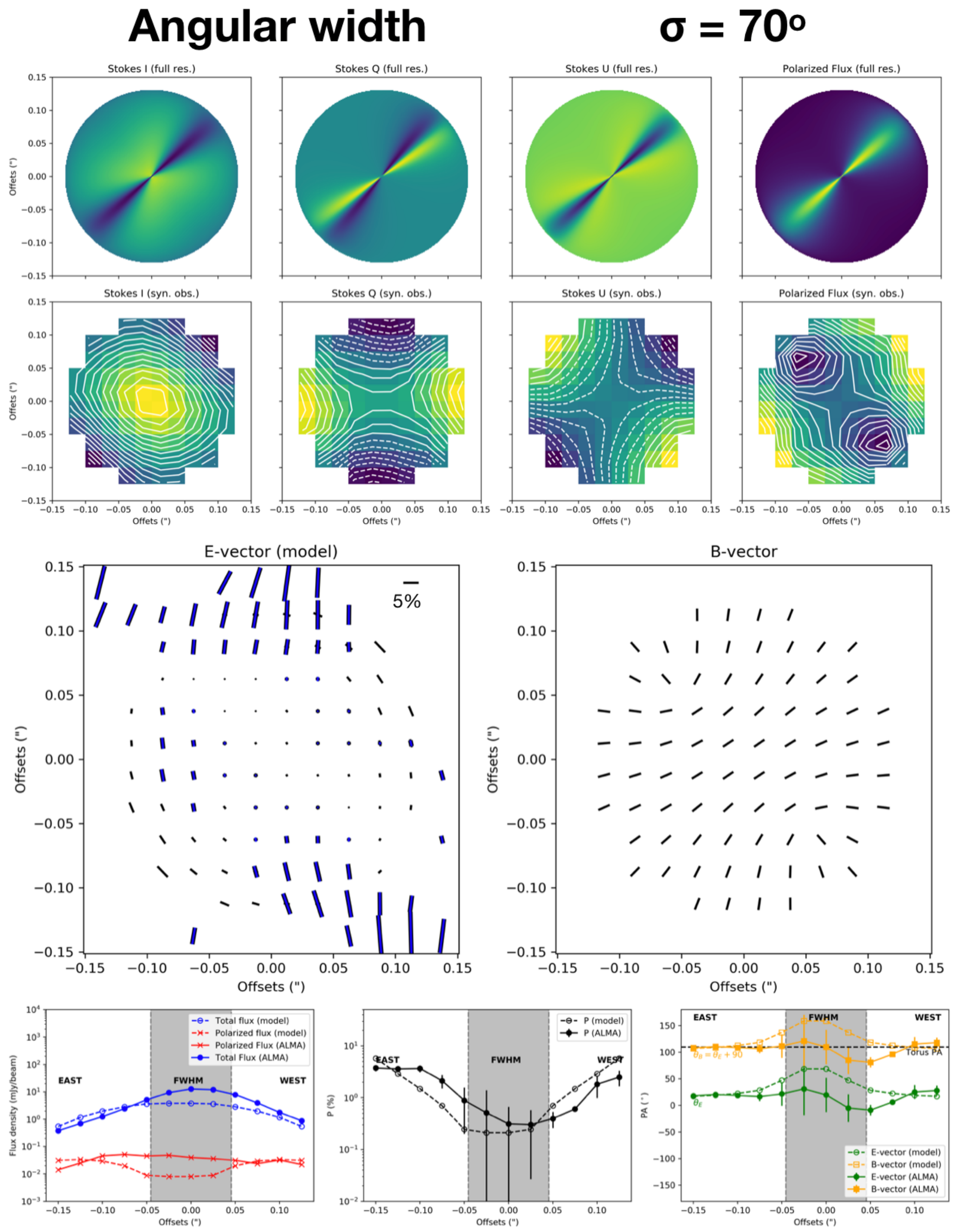}
\caption{Magnetic field model for an angular width of $\sigma = 70^{\circ}$. Same figures as Fig. \ref{fig:fig7}. For $\sigma > 60^{\circ}$, the profile of polarization across the equatorial plane is deeper than the observations.
}
\label{fig:fig10}
\epsscale{2.}
\end{figure*}

\begin{figure}[ht!]
\includegraphics[angle=0,scale=0.80]{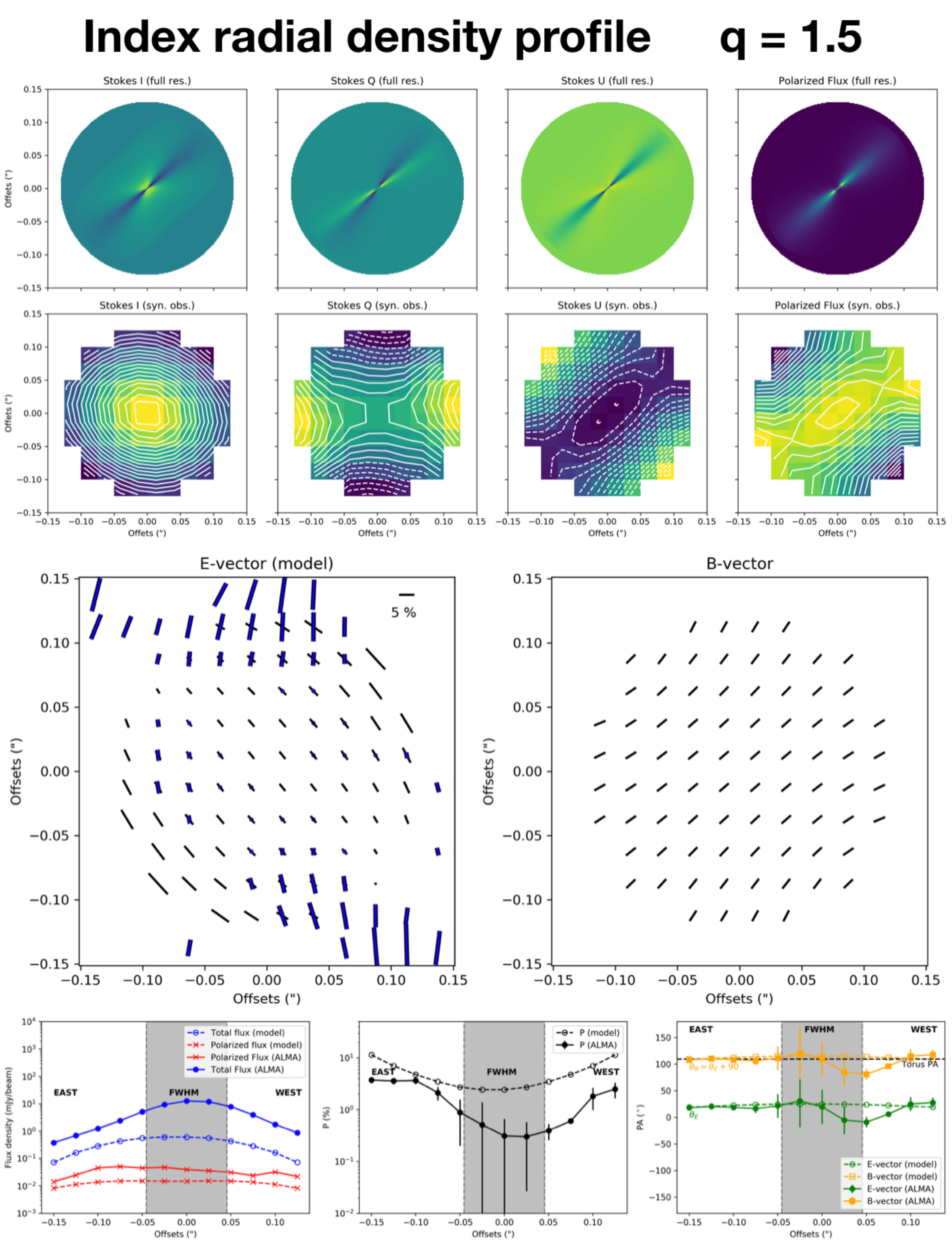}
\caption{Magnetic field model for an index radial profile, $q$, of $1.5$. Same figures as Fig. \ref{fig:fig7}. For $q>1.0$, the torus is very compact and polarization of the core can be detected.
}
\label{fig:fig11}
\epsscale{2.}
\end{figure}

\end{document}